\documentclass[epj]{svjour}
\usepackage{cite}
\usepackage{graphics}
\usepackage{color}
\newcommand{\bm}[1]{\mbox{\boldmath $#1$}}

\newcommand{\dsbabar}{$D_{s0}^*(2317)$}
\newcommand{\dbelle}{$D_{0}^*$(2300--2400)}
\begin{document}
\title{Multichannel calculation of the very narrow
\bm{D_{s0}^*}(2317) \\
and the very broad \bm{D_0^*}(2300-2400)}
\author{George Rupp\inst{1} \and Eef van Beveren\inst{2}% etc
% \thanks is optional - remove next line if not needed
}                     % Do not remove
\institute{
Centro de F\'{\i}sica das Interac\c{c}\~{o}es Fundamentais, Instituto Superior
T\'{e}cnico, Edif\'{\i}cio Ci\^{e}ncia, P-1049-001 Lisboa, Portugal
\and
Centro de F\'{\i}sica Te\'{o}rica, Departamento de F\'{\i}sica,
Universidade de Coimbra, P-3004-516 Coimbra, Portugal}
\date{Received: date / Revised version: date}
% The correct dates will be entered by Springer
%
\abstract{
The narrow $D_{s0}^{\ast}$(2317) and broad $D_0^{\ast}$(2300--2400) charmed
scalar mesons and their radial excitations are described in a coupled-channel
quark model that also reproduces the properties of the light scalar nonet.
All two-meson channels containing ground-state pseudoscalars and vectors are
included. The parameters are chosen fixed at published values, except for the
overall coupling constant $\lambda$, which is fine-tuned to reproduce the
$D_{s0}^{\ast}$(2317) mass, and a damping constant $\alpha$ for subthreshold
contributions. Variations of $\lambda$ and $D_0^{\ast}$(2300--2400) pole
postions are studied for different $\alpha$ values. Calculated cross sections
for $S$-wave $DK$ and $D\pi$ scattering, as well as resonance pole positions,
are given for the value of $\alpha$ that fits the light scalars. The thus
predicted radially excited state ${D_{s0}^*}'(2850)$, with a width of about 
50 MeV, seems to have been observed already.
 % end of PACS codes
} %end of abstract
\authorrunning{E.~van Beveren and G.~Rupp}
\titlerunning{Multichannel calculation of very narrow $D_{sJ}^*$(2317) and
very broad $D_0^*$(2300-2400)}
\maketitle
The very narrow \dsbabar\ charm-strange scalar meson first observed
\cite{PRL90p242001} three years ago has turned out to be the precursor of a
series of new discoveries in hadron spectroscopy that have breathed new life
into this field.  The surprisingly low mass of the \dsbabar\ itself has given
rise to a flurry of theoretical work and speculations, mostly embracing
non-standard quark configurations (see e.g.\ Ref.\ \cite{PRD74p037501} for a
list of references). Moreover, the very broad charm-nonstrange partner meson
\dbelle\footnote
{We adopt here the designation \dbelle\ instead of the official PDG
nomenclature $D_0^*(2400)$ \cite{PDG2006}, to roughly indicate the two
\cite{PRD69p112002,PLB586p11} observed mass values.}
discovered \cite{PRD69p112002} shortly afterwards further added to the
confusion, as its Breit-Wigner mass seems of the same order as the mass of
the \dsbabar, and perhaps even larger \cite{PLB586p11}. However, the large
width ($\simeq260$ MeV) of the \dbelle\ and the conflicting experimental mass
determinations leave enough room for a possible reconciliation with the
\dsbabar\ mass.

In Ref.~\cite{PRL91p012003}, we described the quasi-bound \dsbabar\ and the
\dbelle\ resonance as
$P$-wave $c\bar{s}$ and $c\bar{n}$ ($n=u,d$) states, respectively, strongly
coupled to the lowest $S$-wave two-meson channel, i.e., $DK$ resp.\ $D\pi$.
The framework of our calculation was a simple coupled-channel model previously
used to fit the $S$-wave $K\pi$ phase shifts and predict the $K_0^*(800)$
(alias $\kappa$) meson \cite{EPJC22p493}. As a result, both the quasi-bound
\dsbabar\ \em below \em \/the $DK$ threshold and the very broad \dbelle\
resonance \em above \em \/the $D\pi$ threshold were roughly reproduced, though
with a too low-lying \dbelle\ pole. Scaling arguments from flavour invariance
later allowed to somewhat improve \cite{MPLA19p1949,AIP756p360,PRD74p037501}
our predictions, with no new parameters. In the present study (also see
Ref.~\cite{HEPPH0606110}), we ameliorate our coupled-channel description, by
including all pseudoscalar-pseudoscalar (PP) and vector-vector (VV) channels,
via a generalisation of our model that has recently been applied with success
\cite{HEPPH0606022} to the whole light scalar nonet. The extension to PP and VV
channels should allow us to make reliable predictions at least up to
$\sim\!3$~GeV.

Inclusion of all (ground-state) PP and VV channels implies that we couple the
scalar $c\bar{s}$ states to $DK$, $D_s\eta$, $D_s\eta'$ in $S$ waves, and to
$D^*K^*$, $D_s^*\phi$ in $S$ as well as $D$ waves, leading to a total number
of 7 meson-meson channels. For the corresponding $c\bar{n}$ states, we need
the coupling to $D\pi$, $D\eta$, $D\eta'$, $D_sK$ in $S$ waves, and to
$D\rho$, $D\omega$, $D_s^*K^*$ in $S$ and $D$ waves, thus totalling 10
channels. For the (bare) confined $c\bar{q}$ states, an infinite
harmonic-oscillator spectrum is taken, as in previous work. These bare states
are then coupled, via the $^{3\!}P_0$ mechanism, to the two-meson channels,
assuming that transitions only occur at a certain distance $r_0$. The
resulting $T$-matrix can be solved in closed form (see
Refs.~\cite{HEPPH0606022,HEPPH0606110} for the formula).

An important and very difficult issue when dealing with coupled channels is
how to treat subthreshold contributions, i.e., the effects of channels that
are kinematically closed. Obviously, one cannot simply neglect channels as
soon as the energy drops below threshold, which would be in gross violation
of analyticity and even of common sense. However, it is also clear that, far
below threshold, only a
\begin{table*}[ht]
\caption[]{Pole positions of $D_0^*$(2300--2400) in MeV.
Parameter $\lambda$ fitted to $D_{s0}^*(2317)$ mass.}
\label{Poles}
\begin{center}
\begin{tabular}{ccccccc}
\hline\noalign{\smallskip}
$\lambda$ (GeV$^{-3/2}$) & $r_0$ (GeV$^{-1}$) &
$\Theta_{\mbox{\scriptsize{PS}}}$ ($^\circ$) & $\alpha$ (GeV$^{-2}$) &
$M_1M_2$ & $D_{s0}^*(2317)$ & $D_0^*$(2300--2400) \\ 
\noalign{\smallskip}\hline\noalign{\smallskip}
2.491 & 3.2 & -13.5 & 0.0 &   PP    & 2.317 & $2186-i109$  \\ 
2.497 & 3.2 & -17.3 & 0.0 &   PP    & 2.317 & $2185-i108$  \\ 
1.468 & 3.2 & -13.5 & 0.0 & PP + VV & 2.317 & $2233-i35.1\!$\\ 
1.469 & 3.2 & -17.3 & 0.0 & PP + VV & 2.317 & $2233-i35.0\!$\\
\noalign{\smallskip}\hline\noalign{\smallskip}
2.700 & 3.2 & -13.5 & 2.0 &   PP    & 2.317 & $2165-i111$  \\ 
2.714 & 3.2 & -17.3 & 2.0 &   PP    & 2.317 & $2163-i110$  \\ 
2.137 & 3.2 & -13.5 & 2.0 & PP + VV & 2.317 & $2205-i66.7\!$\\ 
2.144 & 3.2 & -17.3 & 2.0 & PP + VV & 2.317 & $2203-i66.3\!$\\ 
\noalign{\smallskip}\hline\noalign{\smallskip}
\bf 2.854 & \bf 3.2 & \bf -13.5 & \bf 4.0 &   \bf PP    & 2.317 &
\boldmath $2149-i111$  \\    
2.868 & 3.2 & -17.3 & 4.0 &   PP    & 2.317 & $2147-i110$  \\ 
\bf 2.617 & \bf 3.2 & \bf -13.5 & \bf 4.0 & \bf PP + VV & 2.317 &
\boldmath $2174-i96.4\!$ \\ 
2.629 & 3.2 & -17.3 & 4.0 & PP + VV & 2.317 & $2172-i95.3\!$\\ 
\noalign{\smallskip}\hline\noalign{\smallskip}
2.988 & 3.2 & -13.5 & 6.0 &   PP    & 2.317 & $2135-i108$  \\ 
3.001 & 3.2 & -17.3 & 6.0 &   PP    & 2.317 & $2133-i107$  \\
2.901 & 3.2 & -13.5 & 6.0 & PP + VV & 2.317 & $2145-i105$  \\ 
2.913 & 3.2 & -17.3 & 6.0 & PP + VV & 2.317 & $2143-i104$  \\ 
\noalign{\smallskip}\hline
\end{tabular}
\end{center}
\end{table*}
non-perturbative field-theoretic treatment of the
Dyson-Schwinger type might provide a rigorous description, since constituent
masses are inexorably subject to major self-energy corrections in deeply
bound systems. Evidently, a non-covariant approach like our coupled-channel
Schr\"{o}\-dinger equation cannot account for such effects, despite the use of
relativistic kinematics, as particles are manifestly on-mass-shell. Suppression
of closed channels due to wave functions, which is naturally
included in our Schr\"{o}\-dinger fomalism, empirically turns out to be
insufficient in relativistic systems, as recently observed in the mentioned
application of our model to the light scalars \cite{HEPPH0606022}. Therefore,
we adopt here the same remedy as employed in the latter paper, and also in
many multichannel data analyses, namely the use of subthreshold form factors.
Thus, for closed channels we multiply the squares of the individual channel
couplings that show up in our closed-form $T$-matrix expression by an
exponential $\exp(\alpha k_i^2)$, where $k_i$ is the relativistic channel
momentum (with $\Re e\,k_i^2<0$) and $\alpha$ is a positive parameter,
assumed to be universal. Clearly, this ansatz is not fully analytic either,
namely on top of a threshold, but at least it is continuous there.

Now we proceed by fine-tuning the overall coupling constant $\lambda$
\cite{HEPPH0606110} so as to reproduce the mass of the nowadays firmly
established \dsbabar. We do this for a variety of situations in which not only
the parameter $\alpha$ is chosen at different values, but also a comparison is
made between calculations with only PP channels included, and with VV channels
accounted for as well. Moreover, we also choose two different but both
frequently quoted values for the pseudoscalar mixing angle
$\theta_{\mbox{\scriptsize PS}}$, which introduces slight variations in the
predictions owing to the channels involving an $\eta$ or $\eta'$ meson. In each
case, we determine the pole position of the ground-state scalar $c\bar{n}$
state. In Table~\ref{Poles}, these pole positions are given together with the
values of the parameters $\lambda$, $\alpha$, $\theta_{\mbox{\scriptsize PS}}$,
and $r_0$ (fixed). From the table, we first of all observe that the dependence
of the pole positions on the pseudoscalar mixing angle is indeed very feeble.
Then, we note that the inclusion of the VV channels has a very significant
effect on the \dbelle\ pole positions, especially on the imaginary parts, which
nevertheless becomes smaller for increasing $\alpha$. Also this can be easily
understood, as all VV channels are highly virtual at these pole energies,
so that large values of $\alpha$ lead to a strong suppression of these
channels. Finally, all pole positions come out too low when compared to both
experimental \cite{PRD69p112002,PLB586p11} \dbelle\ masses, even when noticing
that the cross sections corresponding to these poles peak at somewhat higher
masses. For instance, in the case $\alpha=4.0$~GeV$^{-2}$ (boldface in
Table~\ref{Poles}), which was the value used in Ref.~\cite{HEPPH0606022} for
all light scalars, our \dbelle\ cross-section peaks lie at 2.18 GeV (PP) and
2.19 GeV (PP+VV). However, some words of caution are due here. Besides the
mentioned 100 MeV discrepancy between the two central experimental masses,
it should be realised that one cannot just compare our predicted cross sections
for elastic scattering with the Breit-Wigner fits of a very broad resonance
observed in production processes, where other and more pronounced resonances
like e.g.\ the $D_2^*(2460)$ show up as well. The resulting distributions for
the \dbelle\ may be quite different (see e.g. FIG.~2 of
Ref.~\cite{HEPPH0606110}). Of course, pole positions must be the
same in elastic scattering and production, but experiment does not extract
any poles from the data. So better data on the \dbelle\ are definitely needed.

\begin{figure*}[ht]
\begin{tabular}{ll}
\resizebox{0.46\textwidth}{!}{\includegraphics{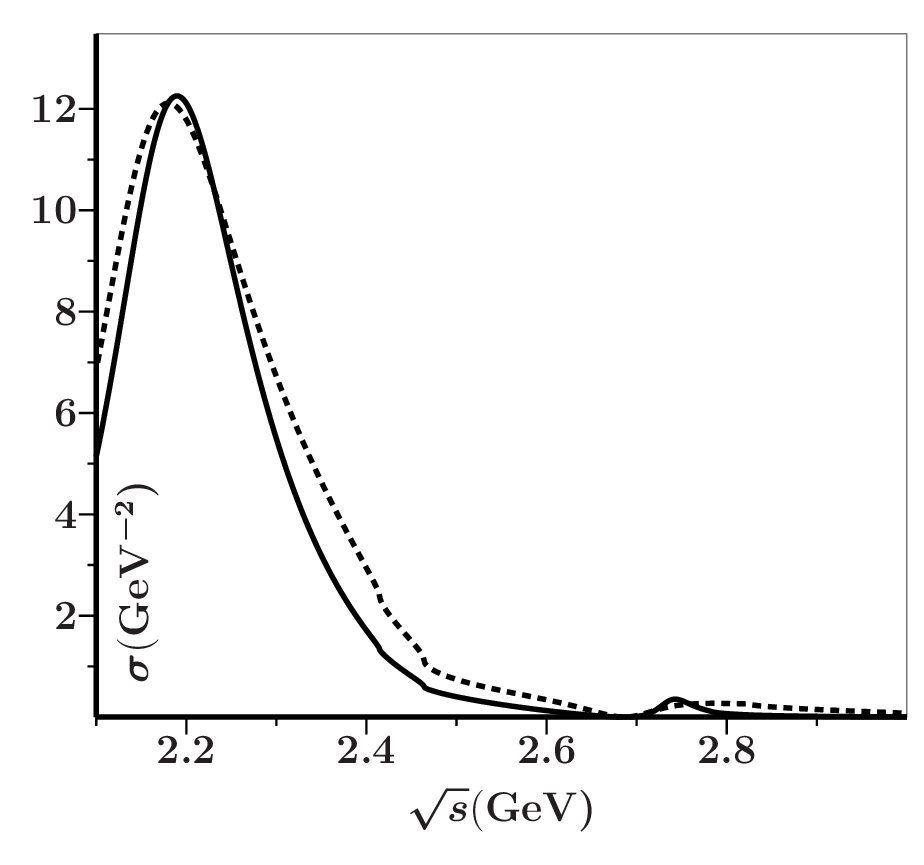}} &
\resizebox{0.46\textwidth}{!}{\includegraphics{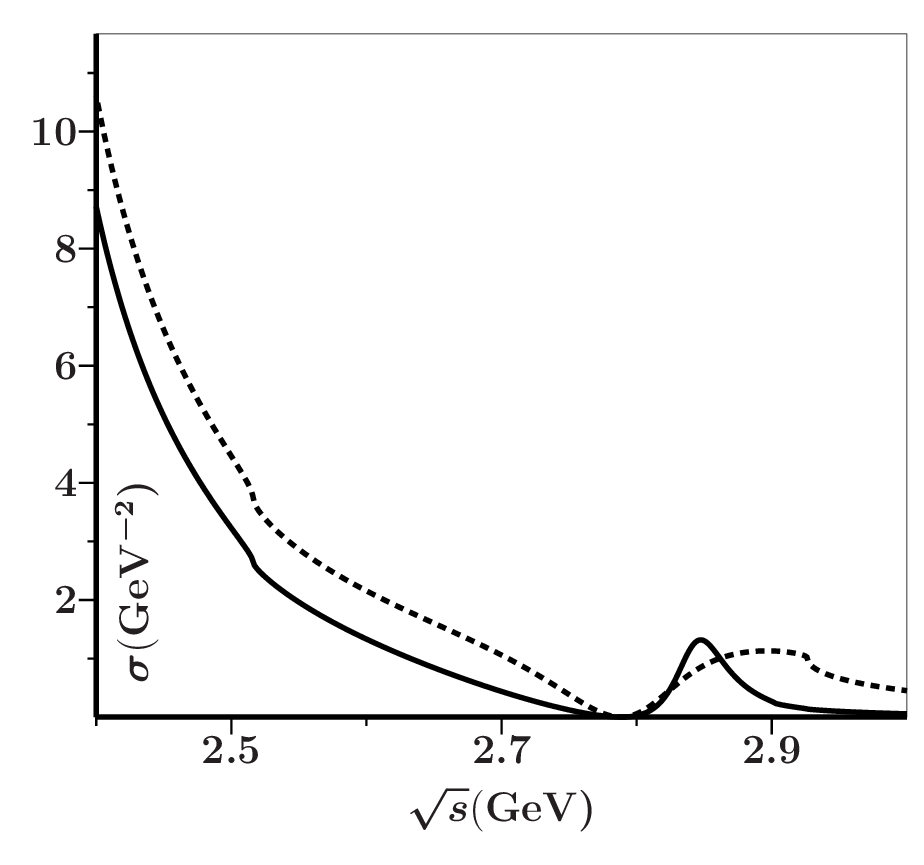}}
\end{tabular}
\caption{$S$-wave $D\pi$ (left) and $DK$ (right) cross sections. Dashed
curves: PP channels only; full curves: PP+VV.}
\label{cross}
\end{figure*}

Focusing now our attention on the case $\alpha=4.0$~GeV$^{-2}$,
 which fits the
light scalars, we compute the elastic $S$-wave $D\pi$ and $DK$ cross sections
up to 3 GeV, which are then plotted in Fig.~\ref{cross}, both for the PP-only
and the full PP+VV cases. Besides the large bump in $D\pi$ due to the \dbelle,
and the steeply falling cross section in $DK$ owing to the \dsbabar\
quasi-bound state, we observe
additional structures at higher energies, which
are more pronounced and narrower when all channels are included. Concretely,
there is a tiny bump in $D\pi$ at about 2.74 GeV, with a peak width of roughly
50 MeV. We find the corresponding pole at $2737-i24.0$ MeV, which can be
traced back, for vanishing $\lambda$, to the first radial excitation of the
bare confinement spectrum at $E=2823$ MeV. Moreover, there is a very broad
pole as well, viz.\ at $2703-i228$ MeV, which is connected to the confinement
ground state, the \dbelle\ being a continuum pole
\cite{PRL91p012003,PRD74p037501}.
In the $DK$ case, there is a clear bump at about 2.85 GeV, with a peak width
of again some 50 MeV. Besides a very broad continuum pole at $2779-i233$ MeV,
there is indeed a narrow pole at $2842-i23.6$ MeV, originating from the bare
confinement state at $E=2925$ MeV. This resonance should thus correspond to
the first radial excitation of the \dsbabar. Quite significantly, a new
$c\bar{s}$ resonance denoted $D_{sJ}(2860)$, with a mass of
$2856.6\pm1.5\pm5.0$ MeV and a width of $48\pm7\pm10$ MeV, was reported
\cite{DsJ2860} by the BABAR collaboration \cite{HEPEX0607082} very shortly
after the presentation of the present results. The observation of the $DK$
decay mode and the non-observation of the $D^*\!K$ mode, as reported by BABAR,
are compatible with the radially excited scalar $c\bar{s}$ state predicted by
us.

In the meantime, three other theoretical papers
\cite{HEPPH0607245,HEPPH0608139,HEPPH0609013} on the $D_{sJ}(2860)$ have
appeared. The first one argues in favour of a $3^-$ assignment, the second one 
supports a radially excited scalar as we do, and the third admits either
option. So experimental confirmation of the $D_{sJ}(2860)$ is needed,
as well as observation of another decay mode.
\begin{acknowledgement}
We are indebted to S.~Tosi for drawing our attention to the $D_{sJ}(2860)$,
right after its first public announcement \cite{DsJ2860}.
This work was supported in part by the {\it Funda\c{c}\~{a}o para a
Ci\^{e}ncia e a Tecnologia} \/of the {\it Minist\'{e}rio da Ci\^{e}ncia,
Tecnologia e Ensino Superior} \/of Portugal, under contract
POCI/FP/63437/2005.
\end{acknowledgement}
\enlargethispage*{20pt}

\end{document}